\documentclass[11pt,a4paper]{article}

\newif\ifjhepstyle
\jhepstylefalse

\ifjhepstyle
	\usepackage{jheppub}
	\usepackage{amsfonts}
	\usepackage{verbatim}
	\usepackage{float}
	\restylefloat{table}
	
\else	\usepackage{verbatim}
	\usepackage{cite}
	\usepackage{setspace}
	\usepackage[top=2.45cm, bottom=2.5cm, left=2.45cm, right=2.45cm]{geometry}
	\usepackage{amsmath,amsfonts,amssymb}
	\usepackage[colorlinks=true
	,urlcolor=blue
	,anchorcolor=blue
	,citecolor=blue
	,filecolor=blue
	,linkcolor=blue
	,menucolor=blue
	,linktoc=page
	]{hyperref}
	\usepackage{float}
	\restylefloat{table}
	
	\numberwithin{equation}{section}
\fi

\newcommand{\ThisIsTheTitle}{The sky is the limit: free boundary conditions\\ in AdS$_3$ Chern-Simons theory}
\newcommand{\ThisIsAuthorOne}{Luis Apolo}

\newcommand{\ThisIsAuthorTwo}{Bo Sundborg}

\newcommand{\ThisIsTheAffiliation}{Department of Physics \& The Oskar Klein Centre, \\
Stockholm University, AlbaNova University Centre, SE-106 91 Stockholm, Sweden}

\newcommand{\ThisIsTheAbstract}{We test the effects of new diffeomorphism invariant boundary terms in SL(2,R)$\times$SL(2,R) Chern-Simons theory. The gravitational interpretation corresponds to free AdS$_3$ boundary conditions, without restrictions on the boundary geometry. The boundary theory is the theory of a string in a target AdS$_3$. Its Virasoro conditions can eliminate ghosts. Generalisations to SL(N,R)$\times$SL(N,R) higher spin theories and many other questions are still unexplored.}

\begin{document}

\ifjhepstyle
\maketitle
\flushbottom
\fi

\long\def\symfootnote[#1]#2{\begingroup%
\def\thefootnote{\fnsymbol{footnote}}\footnote[#1]{#2}\endgroup} 

\def\({\left (}
\def\){\right )}
\def\lb{\left [}
\def\rb{\right ]}
\def\lB{\left \{}
\def\rB{\right \}}

\def\Int#1#2{\int \textrm{d}^{#1} x \sqrt{|#2|}}
\def\Bra#1{\left\langle#1\right|} 
\def\Ket#1{\left|#1\right\rangle}
\def\BraKet#1#2{\left\langle#1|#2\right\rangle} 
\def\Vev#1{\left\langle#1\right\rangle}
\def\Vevm#1{\left\langle \Phi |#1| \Phi \right\rangle}\def\bbox{\bar{\Box}}
\def\til#1{\tilde{#1}}
\def\wtil#1{\widetilde{#1}}
\def\ph#1{\phantom{#1}}

\def\ra{\rightarrow}
\def\la{\leftarrow}
\def\lra{\leftrightarrow}
\def\p{\partial}
\def\diff{\mathrm{d}}

\def\sinh{\mathrm{sinh}}
\def\cosh{\mathrm{cosh}}
\def\tanh{\mathrm{tanh}}
\def\coth{\mathrm{coth}}
\def\sech{\mathrm{sech}}
\def\csch{\mathrm{csch}}

\def\a{\alpha}
\def\b{\beta}
\def\g{\gamma}
\def\d{\delta}
\def\e{\epsilon}
\def\ve{\varepsilon}
\def\k{\kappa}
\def\l{\lambda}
\def\n{\nabla}
\def\om{\omega}
\def\s{\sigma}
\def\t{\theta}
\def\z{\zeta}
\def\vp{\varphi}

\def\ss{\Sigma}
\def\dd{\Delta}
\def\gg{\Gamma}
\def\ll{\Lambda}
\def\tt{\Theta}

\def\D{{\cal D}}
\def\F{{\cal F}}
\def\H{{\cal H}}
\def\I{{\cal I}}
\def\J{{\cal J}}
\def\K{{\cal K}}
\def\L{{\cal L}}
\def\O{{\cal O}}
\def\P{{\cal P}}
\def\W{{\cal W}}
\def\X{{\cal X}}
\def\Z{{\cal Z}}

\def\we{\wedge}

\def\zz{\bar z}
\def\xx{\bar x}
\def\xp{x^{+}}
\def\xm{x^{-}}

\def\VirU1{\mathrm{Vir}\otimes\hat{\mathrm{U}}(1)}
\def\VirSL2R{\mathrm{Vir}\otimes\widehat{\mathrm{SL}}(2,\mathbb{R})}
\def\U1{\hat{\mathrm{U}}(1)}
\def\SL2R{\widehat{\mathrm{SL}}(2,\mathbb{R})}
\def\sl2r{\mathrm{SL}(2,\mathbb{R})}
\def\by{\mathrm{BY}}

\def\tr{\mathrm{tr}}

\def\sint{\int_{\ss}}
\def\dsint{\int_{\p\ss}}

\newcommand{\eq}[1]{\begin{align}#1\end{align}}
\newcommand{\eqst}[1]{\begin{align*}#1\end{align*}}

\newcommand{\absq}[1]{{\textstyle\sqrt{|#1|}}}


\unless\ifjhepstyle
\begin{titlepage}
\begin{center}

\ph{.}

\vskip 4 cm

{\Large \bf \ThisIsTheTitle}

\vskip 1 cm

{{\ThisIsAuthorOne} and {\ThisIsAuthorTwo}}

\vskip .5 cm

{\em \ThisIsTheAffiliation}

\end{center}

\vskip 1.25 cm

\begin{abstract}
\noindent \ThisIsTheAbstract
\end{abstract}
\end{titlepage}
\newpage
\fi



\section{The sky}

\subsection{is the limit}

By the sky is meant the imagined surface (at infinity) from which light can reach us. Under certain circumstances this surface is the boundary of spacetime. We do not wish to assume anything about the geometry of the boundary, in other words we impose no prior geometry.

AdS gravity depends on the boundaries. Physicists are often cavalier about boundary conditions: our main interest lies in equations of motion and their consequences. AdS/CFT duality is an exception to this rule since the coupling of boundary operators to bulk degrees of freedom constitutes an important part of the dictionary relating two theories. In an early precursor to this duality Brown and Henneaux \cite{Brown:1986nw} related 2d conformal symmetries to asymptotic symmetries of 3d gravity. Three-dimensional gravity being a theory without local degrees of freedom, boundary conditions are all important in this case. It has been argued to be very closely related (if not defined by) 3d Chern-Simons theory with gauge group identified with the local spacetime symmetry group \cite{Achucarro:1987vz,Witten:1988hc}. For the case of a negative cosmological constant, the symmetry can be written as $SL(2,R)\times SL(2,R)$. 

New asymptotic boundary conditions are potentially important in both gravity and higher spin theory. Notably, the conditions at boundaries of spacetime are not uniquely determined. Several choices are possible. The main requirement is that a unique time evolution can be ascertained, at least locally. Famous examples in the present context are the Brown-Henneaux boundary conditions \cite{Brown:1986nw} which entail specific fall off conditions on components of the metric. This is standard and entirely consistent, even if it involves a choice of coordinates in the bulk and a specific boundary geometry. Since the boundaries are crucial in 3d gravity, and diffeomorphism invariance is a time-honoured principle, it could be worthwhile to formulate boundary conditions in a purely geometric way and without prescribing the boundary geometry. This could be achieved by not imposing any structure on the sky, i.e. by employing free boundary conditions\footnote{For previous work in this direction, see ref.~\cite{Compere:2008us,Compere:2013bya,Troessaert:2013fma,Avery:2013dja,Apolo:2014tua}.}. We studied this approach in~\cite{Apolo:2015fja}, and discuss this work and some further consequences of it here. Because $SL(2,R)\times SL(2,R)$ Chern-Simons theory can be regarded as the simplest 3d $SL(N,R)\times SL(N,R)$ higher spin theory, and because higher spin theories in general can be formulated as generalised Chern-Simons theories (see for instance ref.~\cite{Vasiliev:1999ba}), we believe that our ideas may be useful also in this extended context.

Pure three-dimensional gravity is notorious by not disclosing its microscopic degrees of freedom \cite{Carlip:2005zn,Maloney:2007ud}, and its existence as a quantum theory is in doubt \cite{Martinec:1998wm}. Before settling this question the crucial boundary conditions should be re-examined, because boundary conditions may lead to new degrees of freedom without introducing matter. Indeed, an effective string description of three-dimensional gravity emerges from this examination\footnote{Previous connections to strings include ref.~\cite{Banados:1999gh,Sundborg:2013bya,Kim:2014bga}.}. A generally covariant formulation typically jeopardises unitarity, but the connections to strings is likely to resolve this issue for AdS$_3$ gravity. 

\section{Introduction}

AdS$_3$ Chern-Simons theory has been advocated \cite{Witten:1988hc} as a soluble quantum theory of gravity, and by the discovery of black hole solutions of the classical theory \cite{Banados:1992wn,Banados:1992gq} the stakes were raised: one could envision calculating the entropy of black holes of varying sizes from first principles and study their formation from collisions. This could be feasible because Chern-Simons theory is in a sense a simple theory: it is essentially topological and contains no local degrees of freedom. Indeed, black hole formation from two colliding massless particles has been described exactly in classical $AdS_3$ gravity \cite{Holst:1999tc}.

Furthermore, the same theory is the simplest in an $SL(N,R)\times SL(N,R)$ series of Chern-Simons theories. All the others are simple higher spin theories without matter. A host of questions on conical singularities \cite{Mansson:2000sj,Castro:2011iw}, black holes \cite{Gutperle:2011kf,Ammon:2011nk,Castro:2011fm,Perez:2014pya} and gauge invariances vs. geometry in higher spin gravity find their simplest manifestations in the $SL(2,R)$ case. Spin 2 is in a sense the lowest high spin.

Unfortunately, hopes about this approach to gravity has dwindled \cite{Witten:2007kt,Maloney:2007ud}. There does not seem to be solutions to the physical questions that could be addressed in the theory, and worse, there does not even seem to be a consensus on what the problems are. One obvious problem, the unitarity of the Chern-Simons theory, is probably solved by the construction explained below, when taken over to the quantum theory. The unitarity issue of higher spin versions of the theory could have similar resolutions.

The confusing state of affairs in AdS$_3$ gravity challenges us to reconsider how to approach it. We take the prominence of boundaries as a clue in this effort: the boundary is truly ``where the action is". We adapt to higher spin generalisations by using the Chern-Simons formulation, but we maintain contact with standard metric gravity. The strategy is to use Chern-Simons theory as a theory of 3d gravity classically, test coordinate invariant boundary conditions and derive a boundary theory. In fact, we demonstrate that it is the world-sheet theory of a string propagating in AdS$_3$!

\section{Diffeomorphism invariant boundaries}

\subsection{The Chern-Simons formulation of gravity}

First, a review of the Chern-Simons description. The vector potentials 
\begin{equation}
  A = \( w^a + e^a \)T_a, \hspace{30pt} \bar{A}  = \( w^a - e^a \)\bar{T_a}, \label{se3:vielbeins}
\end{equation}
are the fundamental fields in the Ach\'ucarro-Townsend-Witten gauge theory description of $AdS_3$ gravity. They are expressed in terms of the vielbein $e$ and the dual $\omega$ to the spin connection. $T_a$ and $\bar{T_a}$ represent generators of the two $SL(2,R)$ factors in the $AdS_3$ group. The metric 
\begin{equation}
 g_{\mu\nu}=\frac{1}{4} \tr \left\{ (A - \bar{A})_\mu (A - \bar{A})_\nu\right\} \label{CompositeMetric}
 \end{equation}
is now a composite field. As a consequence there is nothing in gauge theory that forbids locally degenerate metrics. The degeneration locus is generically of codimension 1. Degenerate metrics are potentially important in the interpretation of the theory, and appear naturally in the description of BTZ black holes, as described later. 
The action that reproduces gravity is the difference of integrals over two Chern-Simons three forms:
\begin{equation}
I_{CS;\rm{bulk}}[A,\bar{A}] =  I_{CS}[A] -  I_{CS}[\bar{A}] 
\end{equation}
\begin{equation}
  I_{CS}[A] = \sint \tr \Big (A\we d A + \frac{2}{3} A\we A \we A \Big ).
\end{equation}

Diffeomorphisms can be represented in a special way in the Chern-Simons formulation. They are equivalent on shell to particular gauge transformations 
\begin{equation}
\delta A = - Du, \qquad \delta \bar{A} = - Du
\end{equation}
with parameter
\begin{equation}
u = \rho^a P_a = v^\mu e_\mu^a P_a
\end{equation}
where $P_a$ generates AdS translations and $v^\mu$ parametrises diffeomorphisms.
The map between gauge transformations and diffeomorphisms degenerates if and only if the metric degenerates.

We may ask if there are interesting solutions with degenerate metrics. This question is somewhat out of line with the rest of this presentation, but it is potentially important for interpreting the Chern-Simons formulation. The equations of motion are flatness conditions
$F=0=\bar{F}$ on the field strengths corresponding to the potentials $A$ and $\bar{A}$. The solutions
\begin{equation}
A = 0 = \bar{A},
\end{equation}
certainly give rise to a degenerate metric but they
are a bit trivial, and do not agree with the gravitational boundary conditions we want to impose.

A more interesting example is the BTZ black hole metric in the form
\begin{equation}
d{s^2} =  - {\sinh ^2}\rho {\left[ {{r_ + }dt - {r_ - }d\phi } \right]^2} + {\cosh ^2}\rho {\left[ {{r_ - }dt - {r_ + }d\phi } \right]^2}
\end{equation}
which degenerates at the outer horizon joining two exterior solutions. But the corresponding gauge potential
\begin{equation}
\begin{array}{*{20}{c}}
{{A^0}^ +  =  - \frac{1}{\ell }\left( {{r_ + } \mp {r_ - }} \right)\sinh \rho \left( {\frac{{dt}}{\ell } \pm d\phi } \right)}\\
{{A^{1 \pm }} = \frac{1}{\ell }\left( {{r_ + } \mp {r_ - }} \right)\cosh \rho \left( {\frac{{dt}}{\ell } \pm d\phi } \right)}\\
{{A^{2 \pm }} =  \pm d\rho },
\end{array}
\end{equation}
is perfectly regular. In metric gravity we are supposed to only use this solution away from the coordinate singularity at the horizon, but the Chern-Simons gauge theory description does not discriminate against this solution at the horizon. It is thus more radical than a simple reformulation. If we try to interpret the above solution in a metric formulation we need energy momentum from a domain wall (string) at the horizon for consistency with Einstein's equations. Perhaps such strings are the missing ingredients in ref.~\cite{Maloney:2007ud}.

\subsection{Chern-Simons theory for double gauge groups $G\times G$.}

The special product structure of the $AdS_3$ isometry group $SL(2,R)\times SL(2,R)$ has consequences for the
description of the boundary degrees of freedom which are always associated with Chern-Simons theory. The 
same structure applies to higher spin generalisations $SL(N,R)\times SL(N,R)$ and it makes sense to use it.

The Einstein-Hilbert action can be rewritten in terms of Chern-Simons fields,
\begin{equation}
  \sint \absq{g}\( R + 2 \) = I_{CS}[A] - I_{CS}[\bar{A}] - \dsint \tr \big ( A \we \bar{A} \big ), \label{se3:ehcs}
\end{equation}
leads to the equations of motion
\begin{equation}
  0 = F = d A + A\we A, \hspace{30pt} 0 = \bar{F} = d \bar{A} + \bar{A} \we \bar{A},
\end{equation}
implying that all solutions are locally of the form,
\begin{equation}
  A = g^{-1} d g, \hspace{30pt} \bar{A} = \bar{g}^{-1} d \bar{g}, \label{se3:puregauge}
\end{equation}
They are locally pure gauge, but not necessarily globally, due to topology and boundary conditions.

Thus, there are no local degrees of freedom, but there can be boundary degrees of freedom. Their action can be found to be
a non-chiral Wess-Zumino-Witten action
\begin{equation}
  S_{WZW}[G] = -\frac{k}{4\pi} \left \{ \frac{1}{2} \dsint \eta^{\mu\nu} \tr \big [ (G^{-1} \p_{\mu} G )(G^{-1} \p_{\nu} G ) \big ] + \frac{1}{3} \int_{\ss} \tr \(G^{-1} d G \)^3 \right \}, \label{se3:string}
\end{equation}
with 
\begin{equation}
G = g \bar{g}^{-1}.
\end{equation}

There are many different derivations, typically using boundary equations of motion\footnote{A benchmark reference is ref.~\cite{Coussaert:1995zp}.} to combine two chiral WZW actions into one non-chiral action. The procedure
involves specification of boundary terms breaking boundary diffeomorphisms (by explicit use of a boundary Minkowski metric $\eta_{\mu\nu}$). 

For the purposes of eventually finding a complete boundary quantum description we wish to derive an off shell boundary theory, and we also wish to keep boundary general covariance. The first requirement was satisfied in a derivation by Arcioni, Blau and O'Loughlin \cite{Arcioni:2002vv}. To restore boundary diffeomorphisms ($\eta_{\mu\nu}$)
we just replace $\eta^{\mu\nu} \to \absq{\g} \g^{\mu\nu} $ and let $\g_{\mu\nu}$ be dynamical.

Given solutions to the Chern-Simons equations
\begin{equation}
 A = g^{-1} d g, \hspace{30pt} \bar{A} = \bar{g}^{-1} d \bar{g}, \label{se3:puregauge}
\end{equation}
using a Wess-Zumino-Witten action
\begin{equation}
  W[g] = -\frac{k}{4\pi} \left \{   \frac{1}{2} \dsint \absq{\g} \g^{\mu\nu} \( g^{-1} \p_{\mu} g\) \( g^{-1} \p_{\nu} g\) + \frac{1}{3} \sint \( g^{-1} d g \)^3\right  \}, \label{se3:wzw}
\end{equation}
and the notation $ \gg^{\mu\nu}_{\pm} = \e^{\mu\nu} \pm \absq{\g} \g^{\mu\nu}$, 
the gravitational action \emph{with appropriate boundary terms} becomes
\begin{eqnarray}
 W[g] + W[\bar{g}^{-1}] -  \frac{k}{4\pi} \dsint \gg^{\mu\nu}_{-} \( g^{-1} \p_{\mu} g\) \( \bar{g}^{-1} \p_{\nu} \bar{g} \) =  W[g\bar{g}^{-1}]\\
  =  S_{WZW}[G] (\eta^{\mu\nu} \to \absq{\g} \g^{\mu\nu})
\end{eqnarray}
from a Polyakov-Wiegmann identity. A generally covariant non-chiral WZW action for $
G = g \bar{g}^{-1}$ results.

This agrees with our intent. The Metric gravity $\leftrightarrow$ Chern-Simons formulation is not a completely off shell relation.
Thus, to start with a bulk off shell formulation would be unnecessarily accurate. However, to capture 
physical degrees of freedom in the boundary a boundary path integral is needed and therefore an off shell boundary theory.

One piece is still missing, the form of the appropriate boundary terms.

\subsection{The Chern-Simons route to the string}

Now write the Cherns-Simons formulation with appropriate boundary terms, which do not specify a particular boundary geometry:
\begin{equation}
  S_{CS} = \frac{k}{4\pi} \left \{ I_{CS}[A] - I_{CS}[\bar{A}] + J_{\a,\b}[A,\bar{A}] \right\}, \label{se3:csaction}
\end{equation}
 where $k = 1/4G_N$ and the boundary term $J_{\a,\b}[A,\bar{A}]$ reads\footnote{In the presence of higher rank symmetric tensors, and higher rank gauge groups, it would be natural to include boundary terms of higher order than quadratic.}
\begin{equation}
  J_{\a,\b}[A,\bar{A}] = \( 2\a - 1 \) \dsint \tr (A\we \bar{A}) \pm \frac{\b}{2} \dsint \absq{\g} \g^{\mu\nu}\tr \left[ \( A - \bar{A}\)_{\mu}\(A - \bar{A}\)_{\nu}\right]. 
\end{equation}
Arcioni, Blau and O'Loughlin take
\begin{itemize}
\item $\alpha = 0$ to get local Lorentz invariance at the boundary. 
\item Also $\b = 1 - 2\a = 1$ for a regular action in the metric formalism. 
\item $\gamma_{\mu\nu}=\eta_{\mu\nu}$.
\end{itemize}
We let the 2d ``world-sheet" metric $\gamma_{\mu\nu}$ be arbitrary. We fix no prior geometry.
This differs from previous approaches.

The modified term in the action is
\begin{equation}
\dsint \absq{\g} \g^{\mu\nu}\tr \big [ \( A - \bar{A}\)_{\mu}\(A - \bar{A}\)_{\nu}\big ] =
4 \dsint \absq{\g} \g^{\mu\nu}g_{\mu\nu}
\end{equation}
by equation (\ref{CompositeMetric}).
We now integrate out $\gamma_{\mu\nu}$.
This will favour no boundary geometry, and give two constraints
\begin{equation}
g_{\mu\nu}=\frac{1}{2}\gamma_{\mu\nu}\gamma^{\rho\sigma}g_{\rho\sigma}.
\end{equation}
Compare the string action, $\int \absq{\g} \g^{\mu\nu}\partial_\mu X \partial_\nu X$, with the same dependence on the 2d metric $\gamma_{\mu\nu}$ and only two Virasoro constraints due to Weyl invariance 
($\gamma_{\mu\nu} \to e^{\phi}\gamma_{\mu\nu}$).

Gauge fix $\gamma_{\mu\nu} \to e^{\phi} \eta_{\mu\nu}$, and
vary the action:
\begin{equation}
  \d S_{CS} = \frac{k}{2\pi} \left \{  \sint \tr \( \d A \we F + \d \bar{A} \we \bar{F} \) - 2 \dsint \tr \left ( e_- \d A_+ - e_+ \d \bar{A}_- \right ) \right \},
\end{equation}
A well-defined action principle is obtained for  $\delta A_{+} = \delta \bar{A}_{-}=0$, which together with factorisation $G = g \bar{g}^{-1}$ reproduce the (standard) boundary conditions
\begin{equation}
A_{+} =  \bar{A}_{-}=0,
\end{equation}
but now after a gauge fixing from a generally covariant boundary action.

There are consequences of starting from a geometric action: there are constraints to consider. As alluded to above, they
appear as equations of motion from variation of $\gamma_{\mu\nu}$:
\begin{equation}
0 = \frac{\delta}{\delta \gamma_{\alpha\beta}} \dsint \absq{\g} \g^{\mu\nu}\tr \big [ \( A - \bar{A}\)_{\mu}\(A - \bar{A}\)_{\nu}\big ] = 4 \frac{\delta}{\delta \gamma_{\alpha\beta}} \dsint \absq{\g} \g^{\mu\nu}g_{\mu\nu}.
\end{equation}
Choosing a $\gamma_{\mu\nu} = e^{\phi}\eta_{\mu\nu}$ gauge, and applying the $A_{+} =  \bar{A}_{-}=0$ boundary conditions, two constraints
affect the metric
\begin{equation}
0 = 4g_{--}= \tr \big [ A _{-} A_{-}\big ] ,\quad 4 g_{-+} = -\tr \big [ A _{-} \bar{A}_{+}\big ], \quad 0= 4g_{++}=\tr \big [ \bar{A}_{+} \bar{A}_{+}\big ]. \label{MetricConstraints}
\end{equation}
Inserting the solutions $  A = g^{-1} d g, \ \bar{A} = \bar{g}^{-1} d \bar{g}$
we find
\begin{equation}
0= \tr \big [ g^{-1} \partial_{-}g  g^{-1} \partial_{-}g\big ] = \tr \big [ \partial_{-}g  g^{-1} \partial_{-}g g^{-1}\big ] = 
\tr \big [ J_{-} J_{-}\big ] , \quad 0 = \tr \big [ \bar{J}_{+}  \bar{J}_{+}\big ], \label{CurrentsConstraints}
\end{equation}
where $J_{-}$ and $\bar{J}_{+}$ are chiral conserved SL(2,R) currents.
We recognise the stress tensor components $T_{--}$ and $T_{++}$ in a WZW CFT, and their vanishing is familiar 
from the world sheet string as Virasoro constraints.

Since a string is more or less defined as a 2d CFT with Virasoro constraints, we conclude that we have a string interpretation 
of 3d gravity.

Some important observations:
\begin{itemize}
    \item $g$ and $\bar{g}$ depend on all bulk coordinates, but lose dependence on one chiral coordinate each close to the boundary, 
    due to the $A_{+} =  \bar{A}_{-}=0$ boundary conditions.
    \item $G = g \bar{g}^{-1}$ at the boundary describes the embedding of the string world sheet into $SL(2,R)$ which is geometrically $AdS_3$.
    \item Gauge potentials are \emph{not} simply related to currents at the boundary: 
    $A_{-} = g^{-1} \partial_{-} g \not\equiv \partial_{-}g  g^{-1} =J_{-},$ in contrast to other approaches, e.g. \cite{Coussaert:1995zp}.
    \item For a string in a conformal gauge, the bulk metric is conformally flat at the boundary with the conformal factor determined 
    by the string solution:
    $g_{++}=g_{--}=0, \quad 4g_{-+}=-\tr \big [ g^{-1} \partial_{-}g  \bar{g}^{-1} \partial_{+}\bar{g}\big ] $.
\end{itemize}

\subsection{Metric AdS$_3$ gravity and diffeomorphism invariant boundaries}

Metric gravity seems to be much more difficult to formulate than the Chern-Simons version in geometries with boundaries. 
It has been done for non-compact spacetimes with boundaries at infinity. Asymptotic fall off conditions like
\begin{equation}
g_{\mu\nu}  \equiv h_{\mu\nu} = r^2 h^{(0)}_{\mu\nu} + \O(r^0), \quad g_{r \mu} = \O(r^{-3}), \quad g_{rr} = r^{-2} + \O(r^{-4}),  \label{se2:freebc}
\end{equation}
permit some terms to be dropped, while other conditions follow from requiring finiteness of charges. Here $r$ is a radial coordinate, and $h^{(0)}_{\mu\nu}$ is the metric at the conformal boundary. 
If the boundary metric is fixed, we have standard Brown-Henneaux type boundary conditions. 
If the boundary metric permitted to vary we have \emph{free} boundary conditions. 

To determine whether boundary conditions make sense it is necessary to check the action principle.
Free boundary are only compatible with a well-defined action principle if the variation with respect to the boundary metric 
vanishes \cite{Compere:2008us,Apolo:2014tua}:
\begin{equation}
  0=T^{BY}_{\mu\nu}\equiv -\frac{2}{\absq{h^{(0)}}} \frac{\d S}{\d h^{(0)\mu\nu}} = -\frac{1}{\k} \big ( K_{\mu\nu} - K h_{\mu\nu} + h_{\mu\nu} \big ), \label{se2:bytmunu}
\end{equation}
where $T^{BY}_{\mu\nu}$ is the Brown-York stress tensor and $K_{\mu\nu}$ is the extrinsic curvature of the boundary. Note that we need the correct boundary terms
for this to work.

To compare with the Chern-Simons discussion boundary conditions which do not specify a prior
boundary geometry should be studied. It is no big surprise that free boundary conditions do the job. 
But
\begin{itemize}
  \item The boundary terms come with non-standard coefficients.
  \item The requirement of a vanishing Brown-York tensor $T^{BY}_{\mu\nu}$ for free boundary conditions gives 3 equations rather than 2 expected from Virasoro constraints. 
\end{itemize}

Boundary terms are constrained by demanding a well-defined variational principle of the generalised action
\begin{equation}
  S_{GR} = \frac{1}{\k} \left \{ \frac{1}{2} \sint \absq{g} \(R + 2\) + \a \dsint \absq{h} K \pm \frac{\b}{2} \dsint \absq{\g} \g^{\mu\nu} h_{\mu\nu} \right \}. \label{se2:polyakovgraction}
\end{equation}
This action is not well defined for arbitrary coefficients of the boundary terms\footnote{Note that a conventional term proportional to $\int \absq{h}$ has been replaced by the last term containing the world sheet metric $\gamma_{\mu\nu}$. Varying with respect to $\gamma$ and reinserting in the action reproduces the standard form of the action.}.
However, if $2\a + \b = 1$ the bulk equations of motion and the vanishing of $T^{BY}_{\mu\nu}$ are enough \cite{Apolo:2015fja,Detournay:2014fva}.
The standard gravity action corresponds to $\a=1, \b=-1$, while our Chern-Simons case corresponds to $\a=0, \b=1$.

Given that the Chern-Simons approach leads to two Virasoro constraints\footnote{Vanishing of the traceless part of a 2d stress tensor.} it not surprising that the traceless part of a two dimensional  stress tensor defined from metric gravity vanishes. This tensor is the Brown-York tensor.
It is less clear what causes the vanishing of the trace part in the Chern-Simons picture. In the boundary string we get a vanishing trace of the stress tensor due to Weyl invariance, but there 
is not even an expression for the trace part which can be studied off shell. It can however be shown that the the vanishing follows from the chirality of the group elements $g$ and $\bar{g}$ on the approach to the boundary.

Returning to the more general framework we can choose an arbitrary world-sheet metric on the boundary, modulo topological obstructions. 
The boundary metric is conformal to the world-sheet metric,
\begin{equation}
h^{(0)}_{\mu\nu}=e^{2\psi}\gamma_{\mu\nu},
\end{equation}
and the subleading terms of the bulk metric are constrained by Virasoro constraints. They imply that the 2d metrics in the surfaces orthogonal to the radial direction are conformally flat\footnote{In Fefferman-Graham coordinates \cite{Fefferman:1985ok}.}. The conformal factor is constrained by vanishing Brown-York tensor (or chirality of $g$ and $\bar{g}$), cf. ref.~\cite{Apolo:2014tua}. For a boundary metric with vanishing scalar curvature (and $\gamma_{\mu\nu}=\eta_{\mu\nu}$) the flat wave equation is obeyed, $\partial_{+}\partial_{-}\psi=0$. Such boundary geometries are even insensitive to quantum corrections from the boundary trace anomaly.

It remains to connect the metric gravity picture with the Chern-Simons/string picture. An explicit extension of world-sheet string solutions to corresponding Chern-Simons solutions provides such a connection if it contains an asymptotic region and is formulated in terms of coordinates that obey the appropriate fall off conditions. In the general case this is work in progress \cite{ASS}, but some known facts are reviewed below.

\subsection{The string and the boundary theory of gravity}

We propose a string interpretation of 3d gravity, inspired by the analogous structures we have found in the string and in
3d gravity. First, a rough sketch of what such an interpretation would entail. If we consider spacetime boundaries with vanishing curvature scalar, the trace anomaly will not ruin conformal invariance and we get the entries without question marks in Table \ref{string/gravity}. The last two lines are tempting but much more tentative. So far it seems that one string is enough to describe the geometries (AdS and BTZ with boundary waves) which are typically considered in AdS$_3$ gravity, with the possible exception of solutions with additional $2\pi$ excess angle conical singularities \cite{Mansson:2000sj}. 

\begin{table}[h]
\caption{Strings and AdS$_3$ Gravity}
\begin{center}
\begin{tabular}{l l}
\hline
String & Gravity \\ \hline
world sheet conformal field theory & boundary CFT \\
$\to$ vanishing stress tensor trace & $\to$ vanishing Brown-York tensor trace \\
2d diffeomorphism & free boundary conditions\\
$\to$ Virasoro constraints &  $\to$ vanishing Brown-York tensor \\ \hline
target space AdS$_3$ metric & boundary action \\
antisymmetric tensor & bulk (Chern-Simons) action \\ \hline
multiple strings? & multiple boundaries? \\
string interactions? & boundary topology change? \\ \hline
\end{tabular}
\end{center}
\label{string/gravity}
\end{table}
More concretely any string solution in terms of $g$ and $\bar{g}$ can be be extended to Chern-Simons solution in several ways. A simple choice is obtained by introducing a radial coordinate $\rho$ and taking 
$g \to g M(\rho)$ and $\bar{g} \to M(\rho)^{-1}\bar{g}$, which solve the equations of motion and the Virasoro constraints for any $\rho$ 
(cf. equation \ref{MetricConstraints}). The conformal factor is proportional to $\tr \big [ g^{-1} \partial_{-}g  \bar{g}^{-1} \partial_{+}\bar{g}\big ] $, and then depends on $\rho$. This is a proof of concept: string solutions generate 3d geometries. These geometries are however not automatically in a useful form  and even those that are in such form are in an unfamiliar conformal gauge, not immediately recognisable in terms of standard parametrisations like that of Ba\~ nados \cite{Banados:1998gg}. Still the asymptotic charges can be evaluated 
\cite{Troessaert:2013fma,Apolo:2014tua}.

\section{Comments and outlook}

We have associated a boundary string theory to AdS$_3$ gravity. The Virasoro algebra that comes with string theory plays the role of gauge generators and constraints, in contrast to the role played by the Brown-Henneaux Virasoro algebra, which is more familiar in the gravitational context. This role prohibits any classical central extension like the one of Brown and Henneaux\cite{Brown:1986nw}, and one may ask how our result connects to theirs. In fact, their algebra should not be expected to be apparent until the gauge freedom has been fixed and the constraint solved. A kind of light-cone gauge can be used for this purpose\cite{Sundborg:2013bya}.

Strings in AdS$_3$ have an interesting zoo of solutions and we already have simple examples \cite{Apolo:2015fja} of a degeneracy: several string solutions map to the same 3d geometry. It would be extremely interesting to find more examples of this behaviour. The typical situation in 3d gravity, where there are too few states compared to the Bekenstein-Hawking entropy could be improved. One expression of this state of affairs is the belief that gravitation is a macroscopic theory and not a microscopic one. Could it be enough to have the Chern-Simons formulation and the resulting string as a microscopic description?

A next step could be to quantise the boundary theory. Since the string in AdS$_3$ is known, the quantum theory is almost within reach. However, we do not have a completely off shell derivation, since the bulk equations of motion have been used in our approach. The precise quantum theory could receive corrections. Furthermore, unless we add extra degrees of freedom we do not have a critical string, which is the closest quantum theory that is known \cite{Maldacena:2000hw}. If we were extremely lucky, these caveats would cancel, and we could extract information on quantum properties directly.

Finally, a few words about the potential applications to higher spin theory. We have argued that 3d gravity is effectively a simple example of such a theory, at least in the Chern-Simons formulation. All the questions that plague 3d gravity are then likely to show up and require a solution. The main higher spin role of our work is to prepare for such a discussion. In practice, one might ask how our approach can be extended. We have given some clues. Higher symmetric tensors can be included in higher order boundary terms\footnote{A discussion of some of the likely terms necessary for such extensions were studied in \cite{Apolo:2015zxh}.} in $SL(N,R)\times SL(N,R)$ theories, and we expect higher order constraints of $W$ algebra type, which would lead to $W$ strings. All of this remains to be worked out.

\bibliographystyle{utphys2}
\bibliography{3dgravity}

\end{document}